# Gamifying the Escape from the Engineering Method Prison

An Innovative Board Game to Teach the Essence Theory to Future Project Managers and Software Engineers

Kai-Kristian Kemell [https://orcid.org/0000-0002-0225-4560], Juhani Risku, Arthur Evensen, Pekka Abrahamsson [https://orcid.org/0000-0002-4360-2226]
Faculty of Information Technology
University of Jyväskylä (JYU)
Jyväskylä, Finland
{kai-kristian.o.kemell, juhani.risku, arthur.n.evensen, pekka.abrahamsson}@jyu.fi

Aleksander Madsen Dahl, Lars Henrik Grytten, Agata Jedryszek, Petter Rostrup
Faculty of Computer Science
Norwegian University of Science and Technology (NTNU)
Trondheim, Norway
{aleksamd, larshg, agataaj, petternr}@stud.ntnu.no

Anh Nguyen-Duc
Department of Business and IT
University of Southeast Norway
Bø I Telemark, Norway
Anh.Nguyen.duc@usn.no

*Abstract*—Software Engineering is an engineering discipline but lacks a solid theoretical foundation. One effort in remedying this situation has been the SEMAT Essence specification. Essence consists of a language for modeling Software Engineering (SE) practices and methods and a kernel containing what its authors describe as being elements that are present in every software development project. In practice, it is a method agnostic project management tool for SE Projects. Using the language of the specification, Essence can be used to model any software development method or practice. Thus, the specification can potentially be applied to any software development context, making it a powerful tool. However, due to the manual work and the learning process involved in modeling practices with Essence, its initial adoption can be tasking for development teams. Due to the importance of project management in SE projects, new project management tools such as Essence are valuable, and facilitating their adoption is consequently important. To tackle this issue in the case of Essence, we present a game-based approach to teaching the use Essence. In this paper, we gamify the learning process by means of an innovative board game. The game is empirically validated in a study involving students from the IT faculty of University of Jyväskylä (n=61). Based on the results, we report the effectiveness of the game-based approach to teaching both Essence and SE project work.

1 INTRODUCTION

Software Engineering (SE) as a discipline is generally seen as lacking in general theories [6] [8]. Practitioners on the field employ a multitude of different SE methods and variations of the more common methods [8], while especially software startups commonly still work with purely ad hoc methods or various combination of mainly Lean and Agile practices [14]. While tackling the situation through the creation of a universal, context-independent software development methodology that suits every SE endeavor might be the ideal solution, this line of action has seen little success so far as is evident from the amount of various methods and practices being employed on the field. One recent effort to address this situation has been *the Essence Theory of Software Engineering* (Essence from here on out), proposed by the SEMAT initiative [8] [19]. Instead of aiming to be a one-size-fits-all SE method, the Essence specification is a modular framework that can instead be used to support the use of the various existing SE methods and practices [8].

Essence is built on the philosophy that methods are not supposed to be exclusive or monolithic by nature. Instead, it would be ideal if practitioners always sought to employ the methods and practices best suited for each SE context individually. In this context, [7] also refer to what they call *method prisons*. Method prison, they argue, is a situation where an organization is locked into using one or several specific method(s), regardless of whether they fit the current SE context of the organization. They consider this to be the normal state of an IT organization.

They posit that this is a result of methods being treated as being monolithic and exclusive, whereas there is actually nothing preventing practitioners from combining and modifying them as they wish. They have intended Essence to be a solution to method prisons by supporting the modification, combination, and tailoring of methods and practices to fit any possible SE context. This view on SE methods and practices proposed by Essence could potentially serve to improve the quality of SE work of practitioner organizations, and warrants studies looking into it. Acting in line with this view of SE methods and practices, however, requires lots of work, reflecting, and planning from the would-be users of Essence.

Being a new tool, Essence has yet to see widespread adoption among practitioners, although it has recently gained some more traction in the academia [20]. One reason for the relatively low practitioner interest is possibly the lack of tools to help implement it, as well as the failure of its would-be users to see its full potential [6]. Due to the modular nature of Essence, its full potential is not realized until it is tailored by its would-be users to suit their specific SE context. This may make it seem less attractive to potential users at a quick glance. Furthermore, learning Essence is not a quick process [15] and may necessitate the taking on a new perspective on the nature of SE methods and practices, which can deter potential users from exploring it.

Acknowledging the perceived difficulty of adopting Essence, the creators of the specification, as well as other individuals interested in it, have made efforts to facilitate the adoption and use of Essence. Some academic studies and other publications have proposed tools to aid in the implementation of the specification in practice (e.g. [6]). In this paper, we chose to tackle the adoption problem by means of gamifying SE project work and the use of Essence by means of a board game.

Although gamification as a concept is relatively new, the idea of using games for learning purposes, or the concept of *serious games* is not at all new [2]. In fact, the idea of using games for educational purposes by far predates digital games as a phenomenon, making gamification not at all limited to digital games specifically [2]. Reference [2] defines gamification to be "the use of game design elements in non-game contexts". In this particular case, we speak of gamification in the sense of gamifying the SE endeavor through means of simulation in the form of a board game, as well as the gamification of the adoption of Essence.

In this study, we develop and evaluate *The Essence of Software Development – The Board Game* through an empirical experiment. In the experiment, we observe groups of IT students play the board game and use mixed methods to gather data from the participants, as is discussed further in the fourth section. More specifically, the purpose of this study is to create an educational board game that fulfills the following objectives:

*1) First year SE students should learn the basic concepts of Essence and SE in a fun way*

*2) The board game should teach a method agnostic view of SE, and that methods are modular*

*3) The board game should teach the importance of teamwork and communication in SE project work*

The rest of this paper is structured in the following manner. Sections 2 and 3 discuss Essence and the board game respectively. We then go over the research methods of the study in section 4 and discuss the experiment in detail in section 5. The data from the experiment is analyzed in section 6. In section 7 we discuss our findings and their implications before concluding the article in the 8th and final section.

2 THE ESSENCE THEORY OF SOFTWARE ENGINEERING

As Essence has yet to become a widespread tool in the industry, and is still relatively new, having originally been proposed in 2012 [8], we will briefly describe the specification and its components in this chapter. The specification was proposed by the SEMAT (Software Engineering Method and Theory) community that consists of a number of different practitioner organizations and academic researchers [19]. The specification comprises both what the authors call a kernel, which they claim involves the elements that are present in every SE endeavor, and a language for extending the kernel as needed. The specification is therefore modular in nature and is intended to be modified as needed to fit any potential SE context. For example, extant literature has shown how to describe SCRUM with Essence [13].

The Essence kernel is split into three areas of concern: Customer, Solution, and Endeavor [8]. The core of the Essence kernel consists of seven alphas, which the authors refer to as "[the essential] things to work with" [8]. The seven alphas are elements the authors of the specification posit are present in every SE endeavor. The alphas are complemented by a number of Activity Spaces, or "[the essential] things to do" [8]. Each Activity space may contain one or more Activities, or no Activities at all [12]. Finally, the kernel also includes a third type of element: competencies [12]. The competencies underline the key capabilities required from the team in order to carry out the endeavor [8].

In practice, as the quoted descriptions above underline, the alphas of the specification are the trackable elements to be worked on. For example, one of the alphas in the kernel is simply 'Software System'; the system that is being worked on [12]. The alphas are to be tracked to measure the progress being made on the SE endeavor at hand [12]. For the purpose of tracking the alphas, each alpha is assigned a set of states that are used to determine the progress on each alpha during the SE endeavor. Each state includes a brief, general description of the state, e.g. "Ready: the system (as a whole) has been accepted for deployment in a live environment", as well as state checklists to help gauge whether the particular state has been reached [12].

Aside from the kernel, the Essence specification includes a language that is to be used in extending the kernel as needed [12]. The language contains the syntax for creating further alphas and other specification elements [12]. Akin to e.g. XML, it uses both natural and formal language to describe the specification elements. Most of the content in the kernel, and any context-specific versions of it, consists of context-dependent natural language while formal language is mainly used to structure the content written in natural language, as well as to guide users in writing it. Three levels of conformance are specified for descriptions written using the language, with level three descriptions being automatically trackable and actionable, and level one descriptions being rather freeform in nature. Lower level descriptions are easier to produce but offer less utility when used in conjunction with external tools for Essence.

In extant literature, Essence has been applied to student contexts before. Reference [16] conducted a field study on Essence by using student teams to assess the framework. The student teams were to use the framework in a real SE project undertaken as a part of their studies, and their utilization of the framework was monitored during the process. The authors concluded that, in comparison to the results of the same course from earlier years, the utilization of Essence seemed to make a difference in how well the project teams. Apart from academic literature, practitioner reports on the use of the framework are available online. For instance, the SEMAT community website features, among other things, experience reports from practitioners, e.g. [4].

3 THE ESSENCE OF SOFTWARE DEVELOPMENT – THE BOARD GAME

The Essence of Software Development board game was developed by IT students from the Norwegian University of Science and Technology under the supervision of the more senior authors of this paper. We developed the board game in this fashion to ensure a student-oriented design approach, i.e. by having students develop a game they themselves would like to play. The game is intended to serve as a game-based learning tool for teaching the use of the Essence specification, as well as SE project work on a more general level.

In designing the game, we worked with several goals in mind. First, the game should be aimed at new SE students as an introduction to both SE project work and Essence. Secondly, the game should, in this vein, include some important elements of Essence. We decided to focus on the core philosophy of Essence: its method agnostic approach to SE project work, as well as the idea of methods being modular in the sense that they ought to be combined in a way that best suits each SE endeavor at hand. Additionally, we included the seven alphas of the Essence kernel into the game: opportunity, stakeholders, requirements, software system, work, team, and way of working are all present in the game under the surface, though just as in real life, they are not always visibly present as you play.

Thirdly, the game was to reflect the cooperative nature of SE project work by encouraging team work and communication rather than competition. Past research has established that team work and communication are two of the most important areas of SE project work [10]. Finally, the board game, despite being a game, was to be reasonably realistic in simulating an SE project. The resulting board game simulates in a simplified manner an SE endeavor and has the players assume the roles of the project team members, with one of the players acting as the team leader or, in other words, project manager. The goal of the game is to work as a team to complete an SE project. This is a rather novel design choice for a board game as most such games tend to focus on competition rather than cooperation, with players either winning or losing as individuals. In this board game, on the other hand, the players either win or lose as a team, much like in a real world SE project.

Each player controls a character in the game, each of which has a certain level of soft skills, hard skills, and energy. Soft skills are required to successfully cooperate on various project tasks, while hard skills are required to finish certain more difficult SE tasks at a high enough level. Energy, on the other hand, is the main resource in the game, spent on various actions and completing tasks in the project. These attributes of each character can be influenced by various events and items as the game goes on. For example, installing a coffee machine in the office results in everyone having a little bit more energy.

Each game starts with the players drawing a scenario card which dictates the nature of the project being worked on. For example, the players might work on a mobile game commissioned by an external client. The simulated SE endeavor then proceeds iteratively, with each iteration marking an arbitrary period of time. The amount of iterations each game takes is pre-determined by the scenario chosen for each game.

In order to finish the project, the players must work on various SE tasks. The number of tasks that are to be completed is denoted by the scenario drawn at the start of each game. The tasks in the game are split into front-end, back-end and architecture tasks. These are also departments physically present on the game board, along with the testing department. Each character works in one of the department, although players are free to switch departments as they wish during the game, but may only work on the tasks of the department their characters are currently located in. Each finished task, save for architecture tasks, is to be tested before deployment, and untested tasks may result in various risks manifesting.

During each iteration, the players are to cooperate in order to figure out how to best split their available resources between the tasks they must complete. There are no turns and each player is free to act as they wish at any given time during the iteration. While communication is encouraged, it is up to the team leader to make the final decision on what each team member is to work on during each iteration. Once the deadline for the scenario is reached after a certain amount of iterations, the team either wins if all tasks are finished, or loses if any tasks remain unfinished. Though the game is based on iterations, the iterations could just as well be called sprints or phases to account for e.g. a more waterfall-oriented development method.

Essence is present in the game in its method agnostic approach to SE. No method is imposed on the players and they may even choose to use an ad hoc approach to SE should they wish. In line with how Essence encourages combining and mixing various methods, the players are free to choose what methods and practices they employ during the project based on what they consider to be the most beneficial combination. Each practice affects the game in some way, and together the practices can heavily influence the way the game proceeds as they offer various beneficial and less beneficial combinations for the players to explore.

4 RESEARCH METHODOLOGY

This study was conducted as a mixed method study, with a focus on qualitative data. We chose a primarily qualitative approach to this study due to the nature of its research problem which is focused on the subjective experiences of the individuals playing the board game. The data were collected through three separate surveys, one multiple choice exam on SE project work, and written reports delivered by the participants. The underlying philosophical approach for this study is interpretivist, with the study explicitly focusing on the subjective perceptions and experiences of the participants [11]. In addition to contributing to the empirical body of knowledge on engineering in the area of Essence in educational use, drawing from the contribution typology that [14] adapted from [18], this study presents a contribution in the form of guidelines.

This study was carried out through an experiment that was conducted over the course of two successive evenings. The participants were to participate either only on the second evening, or on both evenings. All the participants of the experiment were students from the IT faculty of University of Jyväskylä. More specifically, some were Computer Science majors while others were Information Systems majors. Thus, all participants had some degree of knowledge of SE Engineering project work. On the other hand, all participants were unfamiliar with Essence.

The goal of the experiment was to evaluate whether the board game fulfilled the objectives presented in the introduction. For this purpose, we collected an extensive set of data, both qualitative and quantitative, on the learning experiences and game experiences of the participants involved in the experiment using multiple methods of data collection. The use of a pre-game and post-game survey was adapted from the gamification evaluation process used by [5] while the contents of the post-game survey were adapted from the evaluation criteria of [17]. Furthermore, we followed the general guidelines for planning experiments in SE of [21] in conducting the experiment and planning the data collection.

First, each of the participants filled out a pre-game survey which focused on demographic information, e.g. their age, the year course of the participants, as well as their previous work experience. Then, after the experiment on both days, the participants filled out a largely quantitative post-game survey. The survey was adapted from the evaluation criteria of [17], with some modifications made to the criteria in order for them to better fit into the context of a board game rather than a digital game. The detailed framework can be found in the results chapter of this paper in Table I. The post-game survey was conducted as a Likert five point scale survey, where the choices varied from "strongly disagree" (1) to "strongly agree" (5), with the statements focusing on the learning experience of the participants (e.g. "I learned something new about Software Engineering"), as well as their experience with the board game (e.g. "I had fun playing the Board Game").

In addition to the pre-game and post-game surveys, the students were asked to complete a multiple-choice examination on Software Engineering projects adapted from several public online sources. Finally, all participants were to deliver a written report of two to four pages on their experiences with the board game after the experiment. For the purpose of the data analysis and reporting of the results, we employed the guidelines from [9].

5 THE EXPERIMENT

The study was carried out on by conducting an experiment on two successive evenings, spanning five hours per evening. The participants were only given instructions to arrive at the location of the experiment at the given time and date, and that the experiment was for a scientific study. This was done to avoid having any of the participants familiarize themselves with Essence beforehand, i.e. to gather data as unbiased as possible about their learning. The participants were to either participate on both evenings or only the second evening. The participants were awarded one or two study credits for their participation based on whether they participated on one evening or both evenings. On the first evening, 37 students participated in the experiment, while 61 participated on the second evening, including the 37 that had also been present on the first evening. The protocol was largely the same for both evenings.

*5.1 The First Day*

On the first day, by 16:00 (4 PM), all participants were to arrive at the scene of the experiment. Once all the participants had arrived at the scene, an introductory speech explaining the rules of the experiment was given. In short, they were to participate for the duration of the entire experiment while following any further instructions. While they were allowed to take short breaks to e.g. use the rest room, they were not allowed to leave for longer periods of time. They were then asked to fill out the pre-game survey

After the introduction and the pre-game survey, on the first evening two of the authors asked four students, eight in total, to join each of them in playing a round of the game to demonstrate it to the other participants. The purpose of this demo round was to make it easier for the participants to understand the game. After approximately thirty minutes of demonstration, the participants, save for those who participated in the demonstration, were split into seven groups.

The groups were formed randomly, decided by having the participants draw a piece of paper with a number between one and seven on it from a mug. Once the groups had been formed, each group was assigned one participant who had taken part in the demonstration round. The eighth demonstration round participant was assigned to one of the groups with five rather than six members in it. At approximately 17:00, the groups had been formed and the participants were instructed to play the game in the groups until the end of the experiment.

The authors observed the process, as it is seen in Fig. 1, in a largely passive fashion. The purpose of the observation was primarily to ensure that each group was playing the game and following the rules.

Towards the end of the first experiment day, at 20:30, the participants were offered pizza and were asked to fill out the post-game survey while enjoying it. After completing the survey and jotting their names down on the list of participants, they were free to leave for the evening.

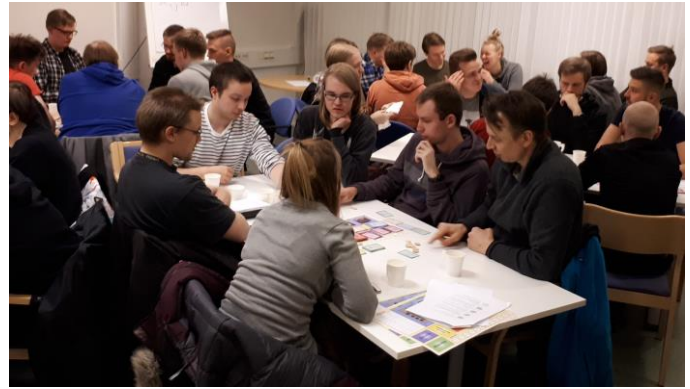

Figure 1 Groups of Participants Playing the Game

*5.2 The Second Day*

The second experiment day was carried out largely in the same fashion. Shortly after 16:00, the participants were once again given an introduction to the experiment. Those that had not participated on the previous day were then asked to fill out the pre-game survey. As over half of the participants had been present on the previous day, no demonstration was given. Instead, the participants were directly split into ten groups in a random fashion, with one group consisting of seven participants and the rest of the groups consisting of six. At 16:30 the participants had been arranged into their respective groups and were asked to play the game until told otherwise.

At 20:10, the participants were asked to start filling out the data collection forms. All participants were asked to fill out the post-game survey, as well as to complete the multiple-choice examination on SE project work. In addition, those participants that had been present on both days were asked to fill out an open-ended survey on the game mechanics of the board game. The purpose of this survey was to collect data that could, in the future, be used to improve the board game, although it was not used in this particular study. At 20:30, the participants were once again offered pizza, and were asked to finish filling out the forms. Once finished with the forms, they were to confirm their attendance and were given instructions for writing their reflective report based on their experiences in the experiment.

## 6 RESULTS

A diverse set of data was gathered from the experiment. The bulk of our findings is based on the quantitative Likert scale survey data from the post-game survey which was conducted following the evaluation criteria of [17], as stated earlier, as well as quantitative data from the multiple-choice examination on SE project work. In addition, these two sets of data are complimented by qualitative data from both the open-ended questions at the end of the post-game survey and the demographic data from the pre-game survey.

The results of the post-game survey are analyzed through the criteria we adapted from [17]. Modifications to the original evaluation criteria of [17] were made to make the framework more applicable to the context of a board game as opposed to a digital game. The main criteria categories of user experience and educational usability were also used to guide the analysis of the data. The criteria, seen in Table I below, were directly converted into statements for the Likert scale post-game survey, the results of which can also be found in the table. The survey results in the table are divided into four columns based on which group of participants the data were collected from. Group A participated in the experiment on both days, while Group B only participated on the second day. This was done to gain a better understanding of how the participants felt about playing the game for longer periods of time.

TABLE I.  EVALUATION CRITERIA AND POST-GAME SURVEY RESULTS

| User experience (UX) | | | | |
|---|---|---|---|---|
| *1. Emotional issues* | *Day 1 (grp A)* | *Day 2 (grp B)* | *Day 2 (grp A)* | *Day 2 (all)* |
| 1.1. Playing the board game motivated me to learn more about Software Engineering | 2,43 | 2,8 | 2,39 | 2,56 |
| 1.2. Playing the board game was fun | 3,43 | 3,04 | 2,33 | 2,62 |
| 1.3. Playing the board game made me want to play more | 2,59 | 2,16 | 1,58 | 1,82 |
| 1.4. This way of learning about SE is exciting | 2,76 | 2,72 | 2,17 | 2,39 |
| 1.5. This way about learning SE is interesting | 2,92 | 3,12 | 2,39 | 2,69 |
| *2. User-centricity/engagement* | *Day 1 (grp A)* | *Day 2 (grp B)* | *Day 2 (grp A)* | *Day 2 (all)* |
| 2.1. The gamification elements enhanced my interest towards studying Software Engineering | 2,76 | 2,92 | 2,44 | 2,64 |
| 2.2. The visual representation of a Software Engineering project enhanced my engagement with the board game | 2,73 | 2,88 | 2,28 | 2,52 |
| 2.3. The interactive way of representing a Software Engineering project enhanced my engagement with the board game | 3,03 | 3,28 | 2,64 | 2,9 |
| 2.4. The textual information about Software Engineering enhanced my engagement with the board game | 2,62 | 2,68 | 2,28 | 2,44 |
| *3. Appeal* | *Day 1 (grp A)* | *Day 2 (grp B)* | *Day 2 (grp A)* | *Day 2 (all)* |
| 3.1. I was interested in playing the board game | 3,57 | 3,64 | 2,25 | 2,82 |
| 3.2. The board game was visually appealing | 2,59 | 2,92 | 2,19 | 2,49 |
| *4. Satisfaction* | *Day 1 (grp A)* | *Day 2 (grp B)* | *Day 2 (grp A)* | *Day 2 (all)* |
| 4.1. The board game experience added fun to the learning opportunity | 3,54 | 3,56 | 2,67 | 3,03 |
| 4.2. This way of learning about Software Engineering is motivating | 3,11 | 2,92 | 2,42 | 2,62 |
| 4.3. I felt a satisfying sense of achievement at some point during the game session | 3,62 | 3,12 | 2,72 | 2,89 |
| 4.4. The board game made me interested in its contents (SE) | 2,95 | 2,96 | 2,42 | 2,64 |
| **Educational usability** | | | | |
| *1. Error recognition, diagnosis and recovery* | *Day 1 (grp A)* | *Day 2 (grp B)* | *Day 2 (grp A)* | *Day 2 (all)* |
| 1.1 The player(s) can make mistakes while playing the board game. I felt like the mistakes I (or we as a team) made were useful learning experiences | 3,16 | 3,08 | 2,56 | 2,77 |
| 1.2 After playing the board game, I feel like I can avoid making similar errors in the future | 2,92 | 2,84 | 2,28 | 2,51 |
| *2. General learning experiences* | *Day 1 (grp A)* | *Day 2 (grp B)* | *Day 2 (grp A)* | *Day 2 (all)* |
| 2.1 Playing the board game resulted in useful learning experiences about Software Engineering | 2,35 | 2,6 | 2,08 | 2,3 |
| 2.2 The contents of the board game (e.g. the vocabulary used) was related to other things I have learned about Software Engineering during my university studies | 3,22 | 3,56 | 2,92 | 3,18 |
| 2.3 The board game taught me new things about Software Engineering | 2,05 | 2,4 | 2,03 | 2,18 |
| 2.4 I feel like the board game was a successful representation of a Software Engineering project | 2,3 | 2,72 | 2,19 | 2,41 |

*6.1 User Experience*

The board game was generally considered to be a positive experience by the participants. The large majority of the participants felt they had both had fun playing the board game and had been interested in doing so. Similarly, the participants generally thought that the board game had added fun to the learning opportunity, and considered a board game to be a motivating way of learning SE. In particular, the participants enjoyed working as a team to win in the game, and some of the participants noted that the social aspect of the gameplay was what they had liked the most about the experience.

Despite having considered the board game experience both fun and interesting, the participants would not have liked to keep playing the game after the duration of the experiment, or even until the very end of it. In their reports and in the open-ended closing questions of the post-game survey, the common sentiment among the participants was that the game was fun for a few rounds, but slowly became less and less interesting as they kept playing. This, many of them added, was a result of the game having little replay value. This can also be seen in Table I when comparing the answers of the participants who participated on both days, i.e. when comparing the responses of group A from the first day to their responses from the second day. Those who participated on both days enjoyed the game less and felt it was less useful on the second day, as evidenced by the averages of almost every survey question. Even the participants who felt the most negative about the game towards the end of the experiment nonetheless typically reported that they had enjoyed the game during the first game round or two.

The participants generally felt that the game became too predictable due to the lack of competitive elements in the board game, and due to the game in general having relatively few random elements in it for a board game. Even more importantly, most participants felt the game was in fact too easy with more than four or five players. This was especially noticeable in the data gathered from the second day of the experiment when the participants were playing in groups of six or seven as opposed to the groups of five on the first experiment day. As the game difficulty did not scale based on the number of players involved in a round, having more players playing the game simply added more resources for the team to use, indeed resulting in the game becoming easier with more players.

As the participants were instructed to keep playing the game until the end of the experiment, some of the groups tackled the problems they felt the game had in terms of game mechanics by establishing house rule. For example, to add an element of competition into the game, one group of participants had one of their members play the role of the "son of the boss". The son of the boss would seemingly be a part of the project team in the game but would seek to sabotage the project from within for his own gain. Some other groups simply lowered the number of players playing the game or imposed restrictions on the amount of resources they had in the game to make the game more difficult and therefore more interesting.

Aside from these game design issues the participants felt the game had, the participants generally reported positive experiences. It is hardly surprising that the participants would not have liked to keep playing the game after already playing it for over four hours in one go, or eight hours on two successive evenings. Given the educational nature of the game, it was not intended to be played for lengthened periods of time for entertainment purposes. After all, once the intended pedagogical goals of the game have been reached, it has served its purpose.

*6.2 Educational Usability*

In evaluating the educational value of the game, we consider teaching both Essence and SE project work as its pedagogical objectives. Though the game is primarily meant to serve as a brief introduction to Essence, the game simulates the process of carrying out an SE project, and consequently is also meant to teach SE project work to students.

The participants largely felt that they had not learned much new about SE while playing the board game, underlining in their qualitative responses that they felt like the game primarily served as a way of revising what they had already learned. Only three respondents agreed with the statement "the board game taught me new things about Software Engineering" in the post-game survey. This sentiment could also be observed through the responses to the post-game survey: 6 participants out of 62 agreed or strongly agreed with the statement "playing the board game taught me new things About Software Engineering." Furthermore, 12 participants out of 62 agreed or strongly agreed with the statement "playing the board game resulted in useful learning experiences about Software Engineering."

While new learning experiences among the participants were seldom reported, 34 out of the 62 participants agreed or strongly agreed with the statement "the contents of the board game (e.g. the vocabulary used) was related to other things I have learned about Software Engineering during my university studies," in addition to 12 participants neither disagreeing nor agreeing with the statement. This suggests that the game does nonetheless successfully teach SE project work in a relevant manner. The participants of the experiment were not limited to first year students, and as a result, largely already had a fair understanding of SE project work. Taking this into account, the lack of new learning experiences is not surprising. It is likely that the game would result in more new learning experiences when played exclusively between first year SE students.

When going into specifics about what they had learned or what they thought the game mainly taught, the responses indicated that the participants felt the game had reinforced their idea of the importance of teamwork in SE project work. Many participants also added that the game emphasized soft skills that they felt are seldom discussed in relation to SE.

Apart from SE project work in general, the board game did not directly teach much about Essence. When asked what they considered the most important in an SE endeavor, based on their experiences with the board game, none of the participants mentioned the kernel or the practices present in the board game. In their written report on the experiment, the participants were also asked to describe Essence in their own words. They were asked to do so without consulting online sources, while at the same time being reminded that the report is not graded and that e.g. "I don't know" is as such a fair answer as well. All of the participants simply wrote that they had no clue as to what Essence was based on the board game. It can nonetheless be argued that the board game did in fact teach the players Essence by conveying the idea of SE methods being modular, along with involving the seven alphas of the Essence kernel, as we will discuss later in this chapter.

*6.3 Objectives of the Board Game*

In the introduction, we defined three objectives for the board game that were evaluated through the experiment. We will now analyze the data directly in relation to these objectives.

*4) First year SE students should learn the basic concepts of Essence and SE in a fun way*

As established in the User Experience subchapter A., the participants nearly universally reported having had fun playing the board game at least for the first one or two rounds, with most of the participants agreeing with the statement "I had fun playing the board game" towards the end of the experiment as well after hours of playing the game. In addition, most participants agreed with the statement "The contents of the board game (e.g. the vocabulary used) was related to other things I have learned about SE during my university studies", which points to the board game successfully capturing the basics of SE project work.

To further gauge whether this goal was reached, we had the participants complete a multiple-choice examination on SE project work after playing the game. The examination was mostly compiled from multiple public online sources, though we added a few additional questions at the end of the survey that were directly related to the contents of the game. However, as we did not have the participants take this examination both before and after the experiment, its results cannot be used make conclusive statements.

The main observation to be made from the multiple-choice examination data is that the majority of the participants passed the examination, as can be seen in Fig. 2. Out of the 61 responses we received in total, 17 were discarded on the basis of being incomplete or otherwise not properly answered, resulting in 45 complete responses. Out of these 45 participants, 34 (75%) would have passed the examination had it been graded, having received more than 50% of the maximum score. The median score was 16 out of 29.

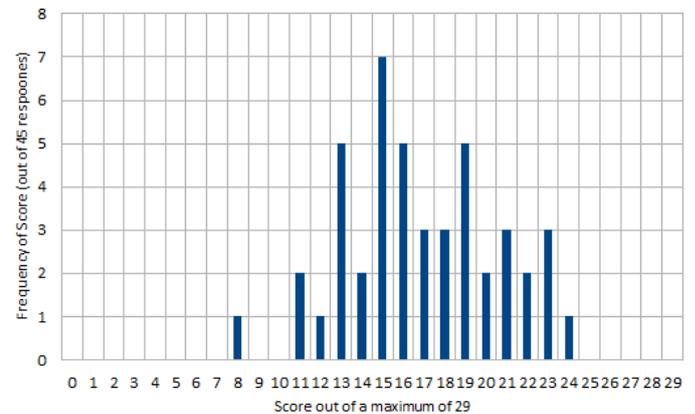

Figure 2    Multiple Choice Examination Results by Score Totals

It is worth noting that there was a possibility of adverse learning while playing the game as well, based on the results of the examination. Being a board game, the game mechanics do result in some generalizations and simplifications of the nature of SE project work, which may be misleading to those with little prior knowledge on the topic. For example, when asked if "the only reason for testing during software development is to mitigate risk at that point in time", 10 respondents out of 45 falsely responded "true". In the context of the board game, that is indeed the only reason to test the software. Furthermore, when asked whether "it's always beneficial to add more developers to a project", in line with how the game became easier the more players (developers) were present, five participants falsely answered "true".

While it is not possible to accurately gauge what effect playing the game may have had on the results of the multiple-choice examination as far as the participant scores go, we nonetheless argue based on our data that this objective was reached. In combination with the multiple-choice examination results, the results of the post-game survey indicate a positive overall result in the context of this objective.

*5) The board game should teach a method agnostic view of SE, and that methods are modular*

This was one of the key principles we followed in designing the game, as was discussed in the third chapter. The participants played the game following the rules as far as the modular use of methods went, and in doing so were introduced to this view on SE methods. More explicit learning experiences in relation to this view on SE methods could certainly be achieved by introducing the players to Essence beforehand, though in this case we chose to not do so to gather as neutral as possible data on what exactly the game taught without outside guidance. Though the participants largely considered Essence to have remained unknown to them after playing the game, we nonetheless argue that this objective was fulfilled through the game mechanics of the game, which pave way for future adoption of Essence among participants.

*6) The board game should teach the importance of teamwork and communication in SE project work*

In response to being asked what they considered important in SE project work based on their experiences with the game, the single most common theme in the responses of the participants was communication and teamwork. One participant, going into more detail, responded that the most important in SE project work was, in their opinion, "an atmosphere that encourages discussion and where one does not have to regret mistakes, as well as communication [in general]". Furthermore, when asked what they had considered to be positive in the game as an open-ended question, a large number of participants mentioned getting to work as a team to have been fun, as well as having enjoyed the social aspect of the game in general. We therefore argue that the third and final objective set for the game was also fulfilled.

## 7 DISCUSSION

Through the experiment, we studied the game-based learning of the Essence specification. Our data indicate that the game-based approach was an enjoyable experience for the participants, and that the board game fulfilled the objectives we outlined in the introduction. In this section, we discuss our findings in relation to teaching Essence, as well as using a board game for educational purposes in the area SE project work.

### 7.1 Implications of the Findings

Extant literature, as well as official SEMAT statements, have suggested that Essence still suffers from a lack of interest among practitioners (e.g. [6] [20]), likely stemming from its resource-intensive adoption and the lack of tools to aid practitioners in adopting it [6]. Past studies in various fields (e.g. [3]) have also shown that game-based learning is a suitable approach. As with any form of teaching, however, the teaching, and in this case the instrument used in it, needs to fit the context and the intended learning goals. We therefore posit that teaching Essence by game-based means is a proposal worth pursuing, serving as a motivation this study. A game-based approach is particularly suitable for this context as the instrument can then be used by other parties to teach Essence and SE in the future.

Analyzing the feedback gathered from the participants on the board game and its game mechanics, the major shortcomings of the game are related to the core game loop which the participants considered to have become too predictable after some rounds, as well as the lack of scaling in the game mechanics. This was an adverse effect of our decision to focus on cooperation and teamwork in designing the game. While the participants enjoyed the social aspect of the game and the cooperation, many of them noted that the lack of competitive elements also made the game less interesting after some time spent playing. To what extent this is to be considered a downside is debatable as the game was not intended to be played for lengthened periods of time. Being an educational game, the game will have already reached its educational objectives after a few rounds. Nonetheless, we did also discover a clear problem we with the game mechanics: the difficulty of the board game presently does not scale based on the number of players. This can make the game too easy, and thus less interesting, when played with a larger group of players.

Aside from these problems the participants reported having had with the game mechanics, the pedagogical side of the game in relation to Essence can also be seen as lacking to some extent based on the data. While the game involves the seven alphas of the Essence kernel, they largely remain under the surface, as discussed in section three. Similarly, though the game is built around the method agnostic nature of Essence that posits that methods and practices should be combined as is seen beneficial in each unique SE context, this is not the focus of the game. Unless the players reflect on this philosophy on their own, they may simply end up playing the game without paying any mind to it. It may thus be beneficial to heighten the role of Essence in the game by e.g. involving the use of the Essence specification language into the gameplay to make the learning experience more purposeful. In its current form, the board game does not directly teach the use of Essence in practice.

Presently, the game is well-suited as a first touch SE project work and project management for new SE students. It is best played for small amounts of time due to the major design decisions behind it which encouraged teamwork and communication at the cost of competitive, replayability-enhancing elements. Our findings indicate that the game successfully: (1) teaches first year Software Engineering students the basic concepts of Essence and Software Engineering in a fun way, (2) teaches a method agnostic view of Software Engineering, and that SE methods are modular, and (3) teaches the importance of teamwork and communication is SE project work.

Putting our findings into a broader perspective, we encourage the use of board games for educational purposes, especially in the context of SE project work. The participants of our experiment reported that they had particularly enjoyed the social aspect of the learning experience and regarded working as a team to beat the game to be an enjoyable activity. Board games offer a chance for students to either learn in a social F2F setting while competing against each other or while collaborating as a team, which is something that students seemed to enjoy based on our results.

### 7.2 Limitations of the Study

The reported results of this study are based on a varied set of data which has some shortcomings. In evaluating what the participants had learned while playing the game, we conducted a multiple-choice exam on SE. However, the data gathered through this exam lacks a point of comparison as it was only gathered after the experiment. It is therefore not possible to accurately determine what exactly was learned from the game and what the participants may have known beforehand. Additionally, though the use of students as subjects for empirical experiments is at times questioned [1], in this case the students were the intended target group of the board game being studied, and thus their use as subjects was well justified.

### 7.3 Recommendations for Future Research

In this paper, we have highlighted some points of improvement in the board game employed in the study. Those interested in developing the game further, or using the game for educational or other purposes, are encouraged to do so as the board game is, as of this publication, available as open source through FigShare. We also have plans to take this board game further so any interested parties are encouraged to contact the authors for possible future cooperation. Though the game examined in this study does succeed in conveying the general philosophy on SE methods behind Essence, it does not concretely teach the use of Essence. This makes it consequently more useful for SE students than practitioners looking to start using Essence. We thus urge those interested in Essence to continue working on tools to help facilitate its adoption. Especially such tools aimed at practitioners are still needed.

### 8 CONCLUSIONS

In this study, we built *The Essence of Software Engineering – The Board Game* to teach the Essence specification and SE project work and demonstrated its effectiveness by means of an empirical experiment. We invited IT students (n=61) to play the board game in an experimental setting and gathered a diverse set of data from the experiment. Based on our findings, we conclude that the board game fulfills the goals set for it. I.e. the board game (1) teaches first year SE students the basic concepts of Essence and SE in a fun way, (2) teaches a method agnostic view of Software Engineering, and that SE methods are modular, and (3) teaches the importance of teamwork and communication is SE project work. On the negative end, our findings indicate that the game has a low replay value and some issues related to game mechanics. Furthermore, the game presently does not teach the use of Essence in practice. To this end, we also discuss possible future improvements to the game and plan on working on it further based on our data. Though the board game is fit to be played as is and is available as such, we will continue to work on the game further and plan to introduce a version with improved replayability through e.g. competitive elements, as well as a heightened role of Essence.

Whereas gamification and serious games are typically discussed primarily in relation to digital games [2], we recommend that board games are also considered for game-based learning purposes in the field of SE. We suggest that future research could investigate the possibility of introducing other board games for teaching SE topics. We also posit that there is still a further need for tools to aid in the adoption of Essence. Due to the central role of project management in the success of SE projects, facilitating the adoption of project management tools is important as well.